\documentclass[a4paper]{article}
\usepackage[psamsfonts]{amssymb}
\usepackage{amsmath}
\usepackage{pstricks}
\usepackage{cite}
\date{}

\author{M. Alimohammadi\footnote{alimohmd@ut.ac.ir}\ \ and A. Ghalee
\\ {\small Department of Physics, University of Tehran,}
\\ {\small North Karegar Ave., Tehran, Iran.}}
\title{Remarks on generalized Gauss-Bonnet dark energy }
\begin{document}
\maketitle
\begin{abstract}
The modified gravity with $F(R,G)$ Lagrangian, $G$ is the
Gauss-Bonnet invariant, is considered. It is shown that the
phantom-divide-line crossing and the deceleration to acceleration
transition generally occur in these models. Our results coincide
with the known results of $f(R)$-gravity and $f(G)$-gravity
models. The contribution of quantum effects to these transitions
is calculated, and it is shown that in some special cases where
there are no transitions in classical level, quantum contributions
can induce transitions. The quantum effects are described via the
account of conformal anomaly.
\end{abstract}
\section{Introduction}
Recent observational data indicate that our universe is currently
in accelerated expansion phase. This is based on the
redshift-distance relationship of type Ia supernovas~\cite{ries},
and many other observations~\cite{sper}. This acceleration is
explained in terms of the so-called dark energy which constitutes
two third of the present universe. There are many candidates which
have been introduced to explain the dark energy: the cosmological
constant, the scalar fields(quintessence or phantom fields),
scalar-tensor theory, k-essence models, etc. Two complete review
articles which have been written in this subject are given
in~\cite{padm}.

Among the interesting features of dark energy, the dynamical
behavior of equation of state parameter $\omega=p/\rho$ is one
which has been studied by many authors. As is well known, the
accelerating universe demands $\omega$ satisfies $\omega<-1/3$.
Also some astrophysical data seem to slightly favor an evolving
dark energy and show a recent $\omega=-1$, the so-called
phantom-divide-line, crossing~\cite{hute}. These observations can
not be explained by single scalar field theories. In quintessence
model, which consists of a normal scalar field, $\omega$ is always
$\omega>-1$, and in phantom model, a scalar field theory with
unusual negative kinetic energy, $\omega$ satisfies $\omega<-1$. A
possible theoretical solution to this problem is to consider the
models known as hybrid models, the models in which there exist
more than one scalar field. One of the famous hybrid model is
quintom model which consists of one quintessence field and one
phantom field. It has been shown that in quintom model with
slowly-varying potentials, the $\omega=-1$ crossing always
occurs~\cite{mohs}.

An alternative approach for explaining the observational data is
to postulate that the gravity is being nowadays modified by some
terms which grow when curvature decrease. The first class of
modified gravities are models known as $f(R)$-gravity, whose
action is a general function $f(R)$ in terms of Ricci scalar $R$,
i.e.
\begin{equation}\label{1}
S=\int d^{4}x\sqrt{-g}{}\hspace{1ex}\left[
\frac{1}{2\kappa^{2}}f(R)+{\cal L}_m \right].
\end{equation}
$\kappa^{2} = 8\pi G$ and ${\cal L}_m$ is the Lagrangian density
of dust-like matter. The possible cosmological applications of
$f(R)$-gravity have been studied in~\cite{nov}. Such theories and
their extensions~\cite{nonv} do pass the local tests and
successfully describe the (almost) $\Lambda CDM$ epoch. For a
review see~\cite{noi}.

Another modification of usual gravity is modified Gauss-Bonnet
theory introduced in~\cite{non,nog}. In this model, the Einstein
action is modified by the function $f(G)$, $G$ being the
Gauss-bonnet (GB) invariant:
\begin{equation}\label{2}
S=\int d^{4}x\sqrt{-g}{}\hspace{1ex}\left[
\frac{1}{2\kappa^{2}}R+f(G)\right] ,
\end{equation}
in which
\begin{equation}\label{3}
G=R^{2}-4R_{\mu\nu}R^{\mu\nu}+R_{\mu\nu\xi\sigma}R^{\mu\nu\xi\sigma}.
\end{equation}
$G$ is topological invariant in four dimensions and may lead to
some interesting cosmological effects in higher dimensional
brane-world approach (for a review, see~\cite{noo}). The
Gauss-Bonnet coupling with scalar field and its contribution in
creation of effective quintessence and phantom era have been
studied in~\cite{nos}. Other aspects of modified GB-gravity, such
as its possibility to describe the inflationary era, transition
from deceleration phase to acceleration phase, crossing the
phantom-divide-line and passing the solar system tests have been
studied in~\cite{noo,cez,leith}.

Recently, a generalization of modified GB-gravity has been
introduced in ~\cite{cez}. The action of this generalized
GB-gravity is:
\begin{equation}\label{4}
S=\int d^{4}x\sqrt{-g}{}\hspace{1ex}\left[ F(R,G)+{\cal L}_m
\right].
\end{equation}
The modified $f(R)$-gravity and modified GB-gravity, i.e. the
actions (\ref{1}) and (\ref{2}), are clearly the special examples
of modified $F(R,G)$-gravity (\ref{4}). The hierarchy problem of
particle physics and the late time cosmology have been studied in
$F(R,G)$ framework in~\cite{cez}. See also \cite{KM}.

The present paper is devoted to the study of some features of
$F(R,G)$-gravity, i.e. the phantom-divide-line crossing and the
possible transition between deceleration and acceleration phases.
We seek the conditions under which these two important crossings
occur, in both classical and quantum levels. It is seen that the
quantum contributions can effectively change the transition
conditions. The quantum effects are described via the the account
of conformal anomaly, reminding about anomaly-driven
inflation~\cite{star}. The contribution of conformal anomaly in
energy conditions and Big Rip of phantom models has been discussed
in~\cite{nol}, and its influence on $\omega=-1$ crossing of
quintessence and phantom model has been studied in~\cite{als}.

The scheme of the paper is as follows. In section 2 we discusses
the Friedman equations of generalized GB-gravity in a spatially
flat Friedman-Robertson-Walker background. Section 3 is devoted to
obtaining the conditions of $\omega=-1$ crossing, and in section 4
the condition for $\ddot{a}<0$ to $\ddot{a}>0$ transition is
derived. It is seen that for special cases where the generalized
GB-gravity is reduced to $f(R)$-gravity or GB-gravity, our results
lead to the known ones for these cases. Finally in section 5, the
contribution of quantum effects on these conditions is obtained
and it is seen that under special initial conditions, these
transitions occur as a result of only quantum effects.

\section{The $F(R,G)$ gravity}

Consider the generalized GB-gravity action ({\ref{4}), with $G$
defined in eq.(\ref{3}). Varying the action (\ref{4}) with respect
to metric $g_{\mu\nu}$ results in ~\cite{cez}:
\begin{equation}\begin{split}\label{5}
&\frac{1}{2}T^{\mu\nu}
+\frac{1}{2}g^{\mu\nu}F(R,G)-2F_{G}(R,G)RR^{\mu\nu}
+4F_{G}(R,G)R^{\mu}_{~\rho}R^{\nu\rho}\\
&-2F_{G}(R,G)R^{\mu\rho\sigma\tau}R^{\nu}_{~\rho\sigma\tau}
-4F_{G}(R,G)R^{\mu\rho\sigma\nu}R_{\rho\sigma}
+2(\nabla^{\mu}\nabla^{\nu}F_{G}(R,G))
R\\&-2g^{\mu\nu}(\nabla^{2}F_{G}(R,G))
R-4(\nabla_{\rho}\nabla^{\mu}F_{G}(R,G))
R^{\nu\rho}-4(\nabla_{\rho}\nabla^{\nu}F_{G}(R,G))R^{\mu\rho}\\
&+4(\nabla^{2}F_{G}(R,G))R^{\mu\nu}
+4g^{\mu\nu}(\nabla_{\rho}\nabla_{\sigma}F_{G}(R,G))R^{\rho\sigma}
-4(\nabla_{\rho}\nabla_{\sigma}F_{G}(R,G))R^{\mu\rho\nu\sigma}\\
&-F_{R}(R,G)R^{\mu\nu}+\nabla^{\mu}\nabla^{\nu}F_{R}(R,G)-
g^{\mu\nu}\nabla^{2}F_{R}(R,G)=0 .
\end{split}\end{equation}
In the above evolution equation, $T_{\mu\nu}$ is energy-momentum
tensor of matter field:
\begin{equation}\label{6}
T_{\mu\nu}=\frac{2}{\sqrt{-g}}\frac{\delta S_{m}}{\delta
g_{\mu\nu}} ,
\end{equation}
and $F_{R}$ and $F_{G}$ are:
\begin{equation}\label{7}
F_{G}(R,G) = \frac{\partial F(R,G)}{\partial G}\hspace{4ex} ,
\hspace{4ex}F_{R}(R,G) = \frac{\partial F(R,G)}{\partial R} .
\end{equation}
A spatially flat Friedman-Robertson-Walker (FRW) space-time in
co-moving coordinates $(t,x,y,z)$ is defined through
\begin{equation}\label{8}
ds^{2}=-dt^{2}+a^2(t)(dx^{2}+dy^{2}+dz^{2}) ,
\end{equation}
where $a(t)$ is the scale factor. For the metric (\ref{8}), the
$(t,t)$-component of (\ref{5}) has the following form:
\begin{equation}\begin{split}\label{9}
 -6H^{2}F_{R}(R,G)=&F(R,G)-RF_{R}(R,G)+6H\dot{F}_{R}(R,G)\\
&+24H^{3}\dot{F}_{G}(R,G)-GF_{G}(R,G) -\rho_{m} .
\end{split}\end{equation}
$H=\dot{a}(t)/a(t)$ is the Hubble parameter, and $\rho_{m}$ is
matter energy density with evolution equation
\begin{equation}\label{10}
\dot{\rho}_{m}+3H(\rho_{m}+p_{m})=0 .
\end{equation}
Here $R$ and $G$ have the following forms:
\begin{equation}\label{11}
R=6(\dot{H}+2H^{2})~\ ~\ ~\  ,~\ ~\ ~\ G=24(H^{2}\dot{H}+H^{4}).
\end{equation}
The sum of $(i,i)$ components of eq.(\ref{5}) for FRW-metric
becomes
\begin{equation}\begin{split}\label{12}
(2\dot{H}+3H^{2})F_{R}(R,G)&=\frac{1}{2}[RF_{R}(R,G)-F(R,G)]-2H\dot{F_{R}}(R,G)
-\ddot{F_{R}}(R,G)\\
&+\frac{1}{2}GF_{G}(R,G)-\frac{G}{3H}\dot{F_{G}}(R,G)
-4H^{2}\ddot{F_{G}}(R,G)-\frac{1}{2}p_{m} .
\end{split}\end{equation}
 Eqs.(\ref{9}) and (\ref{12}) are the Friedman
equations of $F(R,G)$ gravity. Note that the
eqs.(\ref{9}-\ref{12}) are not independent, i.e. the time
derivative of eq.(\ref{9}), using eqs.(\ref{10}) and (\ref{11}),
leads to a linear combination of eqs.(\ref{9}) and (\ref{12}).

\section{The $\omega=-1$ crossing}
In the case of Einstein gravity with scalar fields responsible for
dark energy, the equation of state parameter $\omega=p/\rho$
satisfies $\omega=-1-\frac{2}{3}\frac{\dot{H}}{H^{2}}$. For other
theories, including the $F(R,G)$-gravity, the effective equation
of state parameter $\omega_{\rm{eff}}$ is also defined through
~\cite{non,cez}

\begin{equation}\label{13}
\omega_{\rm{eff}}=\frac{p}{\rho}=-1-\frac{2}{3}\frac{\dot{H}}{H^{2}}.
\end{equation}
So if $H(t)$ has a relative extremum at $t=t_{0}$, the system
crosses $\omega=-1$ line at time $t=t_{0}$.

 Restricting ourselves to $t-t_{0}<<h_{0}^{-1}$, where
 $h_{0}=H(t_{0})$ and $h_{0}^{-1}$ is of order of the age of
 universe, the Hubble parameter can be expanded as
 \begin{equation}\label{14}
H(t)=h_{0}+h_{1}(t-t_{0})^{\alpha}+h_{2}(t-t_{0})^{\alpha+1}+O\left((t-t_0)^{\alpha+2}\right)
,
\end{equation}
in which $\alpha\geq2$ is the order of first non-vanishing
derivative of $H(t)$ at $t=t_{0}$ and
$h_{1}=\frac{1}{\alpha!}H^{(\alpha)}(t_0)$. $H^{(n)}(t_{0})$ is
the $n$-th derivative of $H(t)$ at $t=t_{0}$. The transition from
$\omega>-1$ to $\omega<-1$ regions occurs when $\alpha$ is even
positive integer and $h_{1}>0$. If $h_{1}<0$ , the reverse
transition occurs. Here we seek a solution for the Friedman
equation (\ref{9}), with $\rho_{m}$, $R$ and $G$ given by
eqs.(\ref{10}) and (\ref{11}), when $H(t)$ is expressed by
eq.(\ref{14}). The obtained value of $\alpha$ will determine the
possibility of $\omega=-1$ crossing of $F(R,G)$-gravity theories.

We first rewrite eq.(\ref{9}) as following
\begin{equation}\label{15}
H^{2}F_{R}(R,G)\equiv b(t)\hspace{1ex} ,
\end{equation}
where
\begin{equation}\begin{split}\label{16}
b(t)=-\frac{1}{6}&\left[F(R,G)-RF_{R}(R,G)+6H\dot{R}F_{RR}+6H(\dot{G}+4H^{2}\dot{R})F_{RG}(R,G)\right.\\
&\left.+24H^{3}\dot{G}F_{GG}(R,G) -GF_{G}(R,G)-\rho_{m}\right].
\end{split}\end{equation}
 In eq.(\ref{16}) we use:
\begin{equation}\label{17}
\frac{d}{dt}f(R,G)=\dot{R}f_{R}+\dot{G}f_{G}\hspace{1ex},
\end{equation}
and the subscripts of $F(R,G)$ denote the partial derivatives,
e.g.
\begin{equation}\label{18}
F_{RG}(R,G)=\frac{\partial^2F(R,G)}{\partial R\partial G}
\hspace{1ex}.
\end{equation}
Expanding both sides of eq.(\ref{15}) near $t_{0}\equiv0$, with
$H(t)$ given by eq.(\ref{14}), result in the following relations :
\begin{equation}\label{19}
h_{0}^{2}F_{R}(0)=b(0)\hspace{1ex},
\end{equation}

\begin{equation}\label{20}
h_{0}^{2}(F_{RR}\dot{R}+F_{RG}\dot{G})_{t=0}=\dot{b}(0)\hspace{1ex},
\end{equation}
and
\begin{equation}\begin{split}\label{21}
&\frac{1}{2}h_{0}^{2}\bigr(F_{RRR}\dot{R}^{2}+2F_{RRG}\dot{R}\dot{G}+F_{RR}
\ddot{R}+F_{RGG}\dot{G}^{2}+F_{RG}\ddot{G}\bigr)_{t=0}\\
&+2h_{0}h_{1}F_{R}(0)\delta_{\alpha,2}=\frac{1}{2}\ddot{b}(0)\hspace{1ex}
.
\end{split}\end{equation}
By $F_{R}(0)$ we mean $F_{R}(R,G)|_{t=0}$. Using eqs.(\ref{10})
and (\ref{11}) and expansion (\ref{14}), two relations (\ref{19})
and (\ref{20}) lead to :
\begin{equation}\begin{split}\label{22}
&-h_{0}^{2}F_{R}(0)+\frac{1}{6}F(0)-4h_{0}^{4}F_{G}(0)-\frac{1}{6}\rho_{m}(0)\\
&+12h_{o}h_{1}(F_{RR}+8h_{0}^{2}F_{RG}+
16h_{0}^{4}F_{GG})_{t=0}\delta_{\alpha,2}=0 \hspace{1ex},
\end{split}\end{equation}
and
\begin{equation}\begin{split}\label{23}
&36h_{0}\left[(F_{RR}+8h_{0}^{2}F_{RG}+16h_{0}^{4}F_{GG})h_{2}\right.\\
&\left.+4(F_{RRR}+12h_{0}^{2}F_{RRG}+48h_{0}^{4}F_{RGG}
+64h_{0}^{6}F_{GGG})h_{1}^{2}\right]_{t=0}\delta_{\alpha,2}\\
&+36h_{1}(F_{RR}+8h_{0}^{2}F_{RG}
+16h_{0}^{4}F_{GG})_{t=0}(h_{0}^{2}\delta_{\alpha,2}+h_{0}\delta_{\alpha,3})
+\frac{1}{2}\gamma_{m}h_{0}\rho_{m}(0)=0.
\end{split}\end{equation}
In above equation, $\gamma_{m}$ is defined by
$\gamma_{m}=1+\omega_{m}$ where $\omega_{m}=p_{m}/\rho_{m}$.

As is clear from eq.(\ref{23}), for $\alpha\geq4$ , the only
solution is
\begin{equation}\label{24}
h_{0}=0  \qquad     \text{or}   \qquad     \rho_{m}(0)=0\; ,
\end{equation}
which both are unphysical. In the case $\alpha=3$, eq.(\ref{22})
does not depend on $h_{1}$ and can be used to determine $h_{0}$ in
terms of $\rho_{m}(0)$. Of course its explicit expression can not
be found before the function $F(R,G)$ is known. Note that at
$t=0$, $R(0)=12h_{0}^{2}$ and $G(0)=24h_{0}^{4}$. On the other
hand, the value of $h_{0}$ is also determined from eq.(\ref{10})
in term of $\rho_{m}$ and ${\dot{\rho}}_{m}$ at $t=0$:
\begin{equation}\label{25}
h_{0}=-\frac{{\dot{\rho}}_{m}(0)}{3\gamma_{m}\rho_{m}(0)}.
\end{equation}
The $\alpha=3$ solution exists only if these two different
expressions are equal, which is a very special choice of initial
values. Under these conditions there is no $\omega=-1$ transition.
Except these fine-tuned initial values, the only remaining
solution is $\alpha=2$ which we now discuss it.

 For $\alpha=2$, eqs.(\ref{22}) and (\ref{23}) result in $h_{1}$ and $h_{2}$ ,
 respectively. The parameter $h_{1}$ becomes
  \begin{equation}\label{26}
h_{1}=\frac{6h_{0}^{2}F_{R}+24h_{0}^{4}F_{G}-F+\rho_{m}(0)}{72h_{0}
(F_{RR}+8h_{0}^{2}F_{RG}+16h_{0}^{4}F_{GG})}\biggm|_{R=12h_{0}^{2},~
G=24h_{0}^{4}} \hspace{1ex},
\end{equation}
which is, in general, a non-zero quantity. So $F(R,G)$-gravity can
explain the phantom-divide-line crossing. Depending on the
explicit form of $F(R,G)$, $h_{1}$ can be positive or negative. In
the case of $f(R)$-gravity
\begin{equation}\label{27}
F(R,G)=\frac{1}{2\kappa^{2}}f(R)\hspace{1ex},
\end{equation}
eq.(\ref{26}) leads to
\begin{equation}\label{28}
h_{1}=\frac{6h_{0}^{2}f'(12h_{0}^{2})-f(12h_{0}^{2})+2\kappa^{2}\rho_{m}(0)}{72h_{0}f''(12h_{0}^{2})},
\end{equation}
which is consistent with the result of ~\cite{noi}. In
Gauss-Bonnet gravity
\begin{equation}\label{29}
F(R,G)=\frac{1}{2\kappa^{2}}R + f(G)\hspace{1ex} ,
\end{equation}
eq.(\ref{26}) results in
\begin{equation}\label{30}
h_{1}=\frac{-\frac{3}{\kappa^{2}}h_{0}^{2}-f(24h_{0}^{2})
+24h_{0}^{4}f'(24h_{0}^{4})+\rho_{m}(0)}{1152h_{0}^{5}f''(24h_{0}^4)}\hspace{1ex},
\end{equation}
which is the same as one obtained in ~\cite{non}.

In special case where
\begin{equation}\label{31}
F_{RR}=F_{RG}=F_{GG}\bigr|_{R=12h_0^2,~ G=24h_0^4}=0\hspace{1ex} ,
\end{equation}
eq.(\ref{26}) is not the solution. In this case, eq.(\ref{22})
determines $h_{0}$, which as has been discussed before, is
possible only for specific choice of initial values. Under these
conditions, $h_{1}$ is determined from eq.(\ref{23}) as following
:
\begin{equation}\label{32}
h_{1}=\biggm[-\frac{1}{288}
\frac{\gamma_{m}\rho_{m}(0)}{F_{RRR}+12h_{0}^{2}F_{RRG}+48h_{0}^{4}
F_{RGG}+64h_{0}^{6}F_{GGG}}\biggm]^{1/2}_{R=12h_{0}^{2},~G=24h_{0}^4}.
\end{equation}
It is clear that this solution exist only when the matter energy
density $\rho_{m}(0)$ is not zero.
\section{Deceleration to acceleration transition}
To study the $\ddot{a}<0$ to $\ddot{a}>0$ transition, it must be
noted that $G=24H^{2}(\dot{H}+H^{2})=24H^{2}\ddot{a}/a$. So at the
point of transition $\ddot{a}=0$, one has $G(t_{0}'\equiv0)=0$. We
expand $H(t)$ and $G(t)$ around $t_{0}'=0$ as following :
\begin{equation}\label{33}
H(t)=H_{0}+H_{1}t+H_{2}t^{2}+\cdots ,
\end{equation}
\begin{equation}\label{34}
G(t)=\dot{G}(0)t+\frac{1}{2}\ddot{G}(0)t^{2}+\cdots=G_{1}t+G_{2}t^{2}+\cdots
\hspace{1ex}.
\end{equation}
Since G(0)=0, one finds
\begin{equation}\label{35}
H_{1}=-H_{0}^{2}\hspace{1ex}.
\end{equation}
Using $G=24H^{2}(\dot{H}+H^{2})$ , eqs.(\ref{33}) and (\ref{35})
determine $G_{1}$ and $G_{2}$ as following :
\begin{equation}\begin{split}\label{36}
&G_{1}=48H_{0}^{2}(H_{2}-H_{0}^{3}),\\
&G_{2}=24H_{0}^{2}(5H_{0}^{4}-2H_{0}H_{2}+3H_{3})\hspace{1ex}.
\end{split}\end{equation}
We seek any consistent solution of Friedman equation (\ref{15}),
along with eqs.(\ref{10}) and (\ref{11}), when $H(t)$ and $G(t)$
are given by eqs.(\ref{33}) and (\ref{34}), respectively. Instead
of relations (\ref{19}) and (\ref{20}), here we obtain :
\begin{equation}\label{37}
H_{0}^{2}F_{R}(0)=b(0)\hspace{1ex} ,
\end{equation}
and
\begin{equation}\label{38}
2H_{0}H_{1}F_{R}(0)+H_{0}^{2}(F_{RR}\dot{R}+F_{RG}\dot{G})_{t=0}=\dot{b}(0)
 \hspace{1ex},
\end{equation}
respectively, from them the parameters $H_{2}$ and $H_{3}$ can be
found in terms of $H_{0}$. ( Note that $H_{1}$ is determined by
eq.(\ref{35})). The result for $H_{2}$ is :
\begin{equation}\label{39}
H_{2}=\frac{-F+144H_{0}^{4}F_{RR}+864H_{0}^{6}F_{RG}+1152H_{0}^{8}
F_{GG}+\rho_{m}(0)}{72H_{0}(F_{RR}+8H_{0}^{2}F_{RG}+16H_{0}^{4}F_{GG})}\biggm|_{R=6H_{0}^{2},~G=0}
\hspace{1ex}.
\end{equation}
$H_{3}$ , and other parameters of Hubble rate, can be also found
consistently, which suggests the existence of the transition
between deceleration and acceleration phases of the universe.

In the case of Gauss-Bonnet gravity, i.e. eq.(\ref{29}),
eq(\ref{39}) is reduced to
\begin{equation}\label{40}
H_{2}=H_{0}^{3}-\frac{1}{24f''(0)H_{0}^{3}}\left(\frac{1}{16\kappa^{2}}
+\frac{f(0)}{48H_{0}^{2}}\right)
 \hspace{1ex},
\end{equation}
which is the same as in ~\cite{non}. (Note that there is a little
mistake in eq.(\ref{12}) of ~\cite{non} and the factor 48 of that
equation must be replaced by 24).

For the cases in which eq.(\ref{31}) satisfies, the solution
(\ref{39}) does not exist. In this case, eq.(\ref{37}) results in
\begin{equation}\label{41}
F(R,G)\biggm|_{R=6H_{0}^{2},~G=0}=\rho_{m}(0)\hspace{1ex} ,
\end{equation}
which determines $H_{0}$ in terms of $\rho_{m}(0)$. This is again
held only for special initial values (by initial, we mean at
transition time $t'=0)$. $H_{2}$ is thus found by solving eq.(38),
which results in two lengthy solutions. Note that if
$F\big|_{R=6H_{0}^{2},~G=0}=0$, this solution exists only when
there is no matter energy density at $t'=0$, i.e. $\rho_{m}(0)=0$.
\section{The quantum corrections}
In this section we study the contributions of quantum effects on
transition conditions. For calculating the quantum corrections we
use a standard method in which the interactions are considered
between the quantum particles and classical gravitational field
~\cite{star,nom}. In this context, the renormalization of
effective action leads to some extra terms in the trace of
energy-momentum tensor, which is known as trace/conformal anomaly.
Classically, this tensor is traceless. The extra terms are:
\begin{equation}\label{42}
T=b(F+\frac{2}{3}\square R)+b'G+b''R\hspace{1ex},
\end{equation}
in which F is the square of 4d Weyl tensor
\begin{equation}\label{43}
F=\frac{1}{3}R^{2}-2R_{\mu\nu}R^{\mu\nu}+R_{\mu\nu\alpha\beta}R^{\mu\nu\alpha\beta}\hspace{1ex},
\end{equation}
and $G$ and $R$ are Gauss-Bonnet and Ricci scalars, respectively.
$b,b'$ and $b''$ are given by
\begin{equation}\begin{split}\label{44}
&b=\frac{N+6N_{1/2}+12N_{1}+611N_{2}-8N_{\rm{HD}}}{120(4\pi)^{2}}\hspace{1ex} ,\\
&b'=-\frac{N+11N_{1/2}+62N_{1}+1411N_{2}-28N_{\rm{HD}}}{360(4\pi)^{2}}
,\hspace{1ex}b''=0,
\end{split}\end{equation}
for situations which exist $N$  scalars, $N_{1/2}$ spinors,
$N_{1}$ vector fields, $N_{2}( = 0,$ or 1) gravitons and
$N_{\rm{HD}}$ higher derivative conformal scalars (including
phantoms).

It is worth noting that the calculation which is leaded to
eq.(\ref{42}) is independent of the gravitational part of the
Lagrangian and therefore can cantribute to both the standard and
modified gravities. In fact, this method is based on the
non-minimal coupling of scalar field to gravity, through the
classical Lagrangian density
\begin{equation}\label{n1}
{\cal{L}}=\frac{1}{2}\left\{
-g^{\mu\nu}\varphi_{,\mu}(x)\varphi_{,\nu}(x)-(m^2 +\xi R(x))
\varphi^2(x)\right\},
\end{equation}
which in one-loop level, results in the trace-anomaly
(\ref{42})\cite{star}.

For FRW metric (\ref{8}), eq.(\ref{42}) results in the following
equations for the contribution of conformal anomaly to $\rho$ and
$p$ ~\cite{noa,noo}:
\begin{equation}\begin{split}\label{45}
\rho_{A}=&-\frac{1}{a^{4}}\left\{b'(6a^{4}H^{4}+a^{2}H^{2})\right.\\
&+(\frac{2}{3}b+b'')[a^{4}(-6H\ddot{H}-18H^{2}\dot{H}+3\dot{H}^{2})+6a^{2}H^{2}]\\
&\left.-2b+6b'-3b''\right\} ,
\end{split}\end{equation}
and
\begin{equation}\begin{split}\label{46}
p_{A}=&b'[6H^{4}+8H^{2}\dot{H}+\frac{1}{a^{2}}(4H^{2}+8\dot{H})]\\
&+(\frac{2}{3}b+b'')[-2\dddot{H}-12H\ddot{H}-18H^{2}\dot{H}-9\dot{H}^{2}\\
&+\frac{1}{a^{2}}(2H^{2}+4\dot{H})]-\frac{-2b+6b'-3b''}{3a^{4}}.
\end{split}\end{equation}
The natural way to consider the quantum correction is to add its
contribution to Friedman equations, i.e. to change $\rho_{m}$ in
eq.(\ref{16}) to $\rho_{m}+\rho_{A}$. In this way we can study the
effects of quantum phenomena on $\omega=-1$ and $\ddot{a}=0$
crossings of the universe.

\subsection{Quantum correction to $\omega=-1$ crossing}
In $\omega=-1$ crossing, we take $\alpha=2$ in eq.(\ref{14}) and
consider eq.(\ref{15}) with $\rho_{m}$ replaced by
$\rho_{m}+\rho_{A}$. Eq.(\ref{19}) is then become
\begin{equation}\label{47}
h_{0}^{2} F_{R}(0)=b(0)+\frac{\rho_{A}(0)}{6}\hspace{1ex},
\end{equation}
from which $h_{1}$ is found as follows:
\begin{equation}\label{48}
 h_{1}^{\rm{q.c.}}=\frac{6h_{0}^{2}F_{R}+24h_{0}^{4}F_{G}-F+\rho_{m}(0)-
 6b'h_{0}^{4}-(4b+12b')(h_{0}/
 a_{0})^{2}+(2b-6b')/a_0^4}{72h_{0}
 (F_{RR}+8h_{0}^{2}F_{RG}+16h_{0}^{4}F_{GG}-b/9)}\biggm|_{R=12h_{0}^{2},G=24h_{0}^{4}}.
\end{equation}
In above equation "q.c." stands for " quantum-corrected". As is
obvious from eq.(\ref{48}), in the limits $b\rightarrow 0$ and
$b'\rightarrow0$, one has $h_{1}^{\rm{q.c.}}\rightarrow
h_{1}^{\rm{cl.}}$, which shows the correct behavior of
$h_{1}^{\rm{q.c.}}$. By classical $h_{1},h_{1}^{\rm{cl.}}$, we
mean eq.(\ref{26}). Also note that the quantum correction terms
are much more smaller than the classical terms. This can be seen
by comparing $\rho_{m}(0)$ in numerator of (\ref{48}) with terms
rise from quantum corrections, if we note that :
\begin{equation}\label{49}
 h_{0}^{4}\thicksim \frac{1}{a_{o}^{4}}<<1 \hspace{1ex} .
\end{equation}
Finally it is interesting to compare $h_{1}^{\rm{q.c.}}$ with
$h_{1}^{\rm{cl.}}$ for the cases where eq.(\ref{31}) holds. In
classical level, $h_{1}^{\rm{cl.}}=0$ if $\rho_{m}(0)=0$ (see
eq.(\ref{32})), but here $h_{1}^{\rm{q.c.}}$ is not zero. Using
(\ref{48}), it becomes
\begin{equation}\label{50}
h_{1}^{\rm{q.c.}}=-\frac{6h_{0}^{2}F_{R}+24h_{0}^{4}F_{G}-F-6b'h_{0}^{4}-
(4b+12b')(h_0/a_0)^{2}+(2b-6b')/a_0^4}{8bh_{0}}\biggm|_{R=12h_{0}^{2},~G=24h_{0}^{4}},
\end{equation}
which is a purely quantum mechanical term. In other words, for
$F(R,G)$s which satisfies (31), the classical transition exists
only when a very specific condition is satisfied by initial values
and also when $\rho_{m}\neq 0$ . In fact, in the classical level,
the $\omega=-1$ crossing is a matter-induced transition. But
quantum mechanically, for the same $F(R,G)$, the transition occurs
with or without matter; it is a purely quantum-induced transition.

\subsection{quantum correction of $\ddot{a}= 0$ crossing}
To study the quantum contributions in deceleration-acceleration
transition, we must solve eq.(\ref{37}), when $b(0)$ replaced by
$b(0)+\rho_A(0)/6$. $H(t)$ and $G(t)$ are giveby eqs.(\ref{33})
and (\ref{34}). The final expression of $H_{2}$ is as follows:
\begin{equation}\begin{split}\label{51}
H_{2}^{\rm{q.c.}}=&\left[-F+144H_{0}^{4}F_{RR}+864H_{0}^{6}F_{RG}+1152H_{0}^{8}
F_{GG}+\rho_{m}(0)\right.\\&+(-6b'-14b)H_{0}^{4}
\left.+(-12b'-4b)(H_{0}/a_{0})^2+(-6b'+2b)/a_{0}^{4}\right]\times\\&1/
\left[72H_{0}(F_{RR}+8H_{0}^{2}F_{RG}+16H_{0}^{4}F_{GG}-b/9)\right]_{R=6H_{0}^{2},~G=0}
.
\end{split}\end{equation}
The above equation has correct classical limit (comparing $b=b'=0$
limit of eq.(\ref{51}) with eq.(\ref{39})), and the correction
terms are much smaller than the classical terms. In special case
where
\begin{equation}\label{52}
F=F_{RR}=F_{RG}=F_{GG}\bigr|_{R=6H^2_{0},~ G=0}=0 \hspace{1ex}
\end{equation}
hold, it was shown that the transition solution exists only when
$\rho_{m}(0)=0$ (see the discussion after eq.(\ref{41})). But
here, under the conditions (52), $H_{2}$ in (51) is yet a
solution, without any constraint on $\rho_{m}$ . In other words,
for the cases characterized by (\ref{52}), the classical
transition occurs only when the matter does not exist, but by
considering the quantum effects, it is seen that this transition
occurs in the presence of ordinary matters. This is a purely
quantum effect.

\section{Conclusion}
To summarize, the generalized Gauss-Bonnet dark energy with
Lagrangian (\ref{4}) has been considered in
Friedman-Robertson-Walker background metric. For times near the
$\omega=-1$ crossing time, $H(t)$ can be expressed by (\ref{14}).
It has been shown that the only general solution is one with
$\alpha=2$. The Hubble parameter $h_{1}$ has been obtained in
(\ref{26}), which proves the existence of phantom-divide-line
crossing in $F(R,G)$-gravity models.

For obtaining an $\ddot{a}=0$ crossing solution, we take $H(t)$ as
(\ref{33}), with constraint (\ref{35}), and the parameter $H_{2}$
of Hubble rate has been found in (\ref{39}). This relation shows
the existence of deceleration-acceleration transition in
generalized GB models.

Finally the quantum corrections have been added by considering the
conformal anomalies (\ref{45}) and (\ref{46}). It has been shown
that these corrections modify $h_{1}^{\rm{cl.}}$ (in
eq.(\ref{26})) to $h_{1}^{\rm{q.c.}}$ (in eq.(\ref{48})). In some
special cases, quantum phenomenon is the only reason for
$\omega=-1$ crossing. The same corrections modify
$H_{2}^{\rm{cl.}}$ (in eq.(\ref{39})) to $H_{2}^{\rm{q.c.}}$ (in
eq.(\ref{51})), and again for some special $F(R,G)$s, the
transition from deceleration to acceleration phases is induced by
quantum effects.

{\bf Acknowledgement:} This work was partially supported by the
"center of excellence in structure of matter" of the Department of
Physics of University of Tehran, and also a research grant from
University of Tehran\\ \\


\begin{thebibliography}{9999}

\bibitem{ries} A. G. Riess et al. [Supernova Search Team Collaboration], Astron.
J. \textbf{116} (1998) 1009; S. Perlmutter et al. [Supernova
Cosmology Project Collaboration], Astrophys. J. \textbf{517}
(1999) 565.

\bibitem{sper} D. N. Spergel et al. [WMAP Collaboration], Astrophys. J.
Suppl. \textbf{148} (2003) 175; A. C. S. Readhead et al.,
Astrophys. J. \textbf{609} (2004) 498; J. H. Goldstein et al.,
Astrophys. J. \textbf{599} (2003) 773; M. Tegmark et al. [SDSS
Collaboration], Phys. Rev. D \textbf{69} (2004) 103501.

\bibitem{padm} T. Padmanabhan. Phy. Repts. \textbf{380} (2003) 235; E. J. Copeland, M. Sami and S. Tsujikawa,
Int. J. Mod. Phy. D \textbf{15} (2006) 1753.

\bibitem{hute}D. Huterer and A. Cooray, Phys. Rev. D \textbf{71} (2005) 023506 ; S. Nesserisa
and L. Perivolaropoulos, Phys. Rev. D \textbf{72} (2005) 123519 ;
U. Seljak, A. Slosar and P. McDonald, JCAP \textbf{0610} (2006)
014.

\bibitem{mohs} H. Mohseni Sadjadi and M. Alimohammadi, Phys. Rev. D \textbf{74} (2006) 043506
; M. Alimohammadi and H. Mohseni Sadjadi, Phys. Lett. B
\textbf{648} (2007) 113.

\bibitem{nov}S. Nojiri and S. D. Odintsov, Phys. Rev. D \textbf{68} (2003) 123512;
S. Capozziello, S. Carloni, and A. Troisi, Recent Res. Dev.
Astrophys. \textbf{1} (2003) 625; S. Carroll, V. Duvvuri, M.
Trodden, and M. Turner, Phys. Rev. D \textbf{70} (2004) 043528; V.
Faraoni, Phys. Rev. D \textbf{76} (2007) 127501; Class. Quant.
Grav. \textbf{24} (2007) 3637; T. Sotiriou and S. Liberati, Annals
Phys. \textbf{322} (2007) 935; S. Rahvar and Y. Sobouti, Mod.
Phys. Lett. A \textbf{23} (2008) 1929; O. Bertolami, C. Boehmer,
T. Harko, and F. Lobo, Phys. Rev. D \textbf{75} (2007) 104016; S.
Carloni, A. Troisi, and P. Dunsby, arXiv:0706.0452[gr-qc]; Phys.
Rev. D \textbf{77} (2008) 024024; S. Capozziello and M.
Francaviglia, Gen. Rel. Grav. \textbf{40} (2008) 357; H. Mohseni
Sadjadi, Phys. Rev. D \textbf{76} (2007) 104024; X. Wu and Z. Zhu,
Phys. Lett. B \textbf{660} (2008) 293; S. Capozziello et al, Phys.
Rev. D \textbf{76} (2007) 104019; JCAP \textbf{0808} (2008) 016;
J. Evans, L. Hall, and P. Caillol, Phys. Rev. D \textbf{77} (2008)
083514; S. Tsujikawa, K. Uddin, S. Mizuno, R. Tavakol, and J.
Yokoyama, Phys. Rev. D \textbf{77} (2008) 103009.

\bibitem{nonv} S. A. Appleby and R. A. Battye, arXiv:0705.3199[astro-ph];
L. Pogosian and A. Silvestri, Phys. Rev. D \textbf{77} (2008)
023503; S. Capozziello and S. Tsujikawa, Phys. Rev. D \textbf{77}
(2008) 107501; K. Bamba, S. Nojiri, and S. D. Odintsov, Phys. Rev.
D \textbf{77} (2008) 123532; S. Nojiri and S. D. Odintsov, Phys.
Lett. B \textbf{652} (2007) 343.

\bibitem{noi} S. Nojiri, S. D. Odintsov, Int. J. Geom. Meth. Mod. Phys. \textbf{4} (2007)
115.

\bibitem{non} S. Nojiri and S. D. Odintsov, Phys. Lett. B \textbf{631} (2005)1.

\bibitem{nog} S. Nojiri, S. D. Odintsov, and O. G. Gorbunova, J. Phys. A \textbf{39} (2006)
6627.

\bibitem{noo} S. Nojiri, S. D. Odintsov, and S. Ogushi, Int. J. Mod. Phy. A \textbf{17} (2002)
4809.

\bibitem{nos} S. Nojiri, S. D . Odintsov, and M. Sasaki, Phys. Rev. D \textbf{71} (2005)
123509.

\bibitem{cez} G. Cognola, E. Elizalde, S. Nojiri, S. D. Odintsov, and S. Zerbini, Phys. Rev. D \textbf{73} (2006)
084007.
\bibitem{leith} B. M. Leith, I . P. Neupane, JCAP \textbf{0705} (2007) 019;
S. Nojiri, S. D. Odintsov and P. V. Trelyakov, Phys. Lett. B
\textbf{651} (2007) 224.
\bibitem{KM} T. Koivisto and D. F. Mota, Phys. Lett. B \textbf{644} (2007)
104; Phys. Rev. D \textbf{75} (2007) 023518.

\bibitem{star} A. Starobinsky, Phys. Lett. B \textbf{91} (1980) 99.

\bibitem{nol} S. Nojiri and S. D. Odintsov, Phys. Lett. B \textbf{562} (2003) 147
; S. Nojiri and S. D. Odintsov, Phys. Lett. B \textbf{595} (2004)
1 ; S. Nojiri and S. D. Odintsov, Phys. Rev. D \textbf{70} (2004)
103522.

\bibitem{als} M. Alimohammadi and L. Sadeghian,
JCAP \textbf{0901} (2009) 035.

\bibitem{nom}N. D. Birrell and P. C. W. Davies "{\it Quantum
fields in curved space}" Cambridge University Press, 1986.

\bibitem{noa} S. Nojiri and S. D. Odinstov, Int. J. Mod. Phys. A \textbf{16} (2001)
3273.



\end{thebibliography}
\end{document}